\documentclass[12pt]{article}
\usepackage{amsmath, amsthm}
\usepackage{graphicx}
\usepackage{natbib}
\usepackage{url} 

\usepackage{graphicx,psfrag,epsf}
\usepackage{amsfonts}
\usepackage{color}
\usepackage{algorithm2e}
\usepackage{bm}
\usepackage{multirow}
\usepackage{booktabs} 
\usepackage{hyperref}

\newcommand{\blind}{0}

\def\mc{\mathcal}
\def\mb{\mathbb}

\newtheorem{theorem}{Theorem}
\newtheorem{assumption}{Assumption}

\begin{document}

\def\spacingset#1{\renewcommand{\baselinestretch}%
{#1}\small\normalsize} \spacingset{1}


\if0\blind
{
  \title{\bf Interval censored recursive forests}
  \author{Hunyong Cho\\
    Department of Biostatistics, \\
    University of North Carolina at Chapel Hill,\\
    \\
    Nicholas P. Jewell\\
   Department of Medical Statistics \& Centre for Statistical Methodology,\\
    London School of Hygiene \& Tropical Medicine,\\
    \\
    and \\
    \\
    Michael R. Kosorok\\
    Department of Biostatistics,\\
    Department of Statistics and Operations Research,\\
	University of North Carolina at Chapel Hill}
  \maketitle
} \fi

\if1\blind
{
  \bigskip
  \bigskip
  \bigskip
  \begin{center}
    {\LARGE\bf Interval censored recursive forests}
  \end{center}
  \medskip
} \fi

\bigskip
\begin{abstract}
We propose interval censored recursive forests (ICRF), an iterative tree ensemble method for interval censored survival data. 
This nonparametric regression estimator addresses the splitting bias problem of existing tree-based methods and iteratively updates survival estimates in a self-consistent manner.
Consistent splitting rules are developed for interval censored data, convergence is monitored using out-of-bag samples, and kernel-smoothing is applied. The ICRF is uniformly consistent and displays high prediction accuracy in both simulations and applications to avalanche and national mortality data. An \texttt{R} package \texttt{icrf} is available on CRAN and Supplementary Materials for this article are available online.
\end{abstract}

\noindent%
\vfill

\newpage
\spacingset{1.5} 

\section{Introduction}
\label{s:intro}
Interval censoring is a widely observed censoring mechanism in survival analysis. In interval censored data, the failure time information is given as a form of an interval that is known to contain the failure time. In that sense, right-censored data, where the failure time is either exactly observed or is known to be later than a certain censoring time, is a special case of interval censored data. However, since interval censored data, in its narrow definition, are not given as exact failure times, analysis of such data is often challenging and unique. For instance, while the Kaplan-Meier estimator designed for right-censored data has a closed-form solution, its counterpart for interval censored data, or the non-parametric maximum likelihood estimator (NPMLE), does not have a closed-form solution \citep{huang1997}.

For censored data, tree-based methods have been widely used \citep{zhou2015}. Survival trees recursively partition data into two parts until they form small, homogeneous subgroups (`the terminal nodes') and estimate the marginal survival probabilities for each terminal node (\citet{gordon1985}, \citet{segal1988}, \citet{ciampi1991}, \citet{leblanc1992}, and \citet{leblanc1993}). The partitioning procedure is usually done by exhaustively examining the degree of heterogeneity at all possible cut-offs along every variable and selecting the cut-off that maximizes heterogeneity. Trees stop partitioning when the terminal nodes become smaller than a predefined size or when further splitting does not bring enough reduction in heterogeneity. 

Random survival forests are constructed by averaging a large number of diverse survival trees (\citet{hothorn2004}, \citet{hothorn2005}, \citet{rsf}, and \citet{rist}). Diversity is induced by randomizations such as subsampling, random variable selection for splitting, and random cut-off selection \citep{ert, mench2020}. As a result, random survival forests have reduced variability relative to survival trees. For example, in \cite{ert}'s extremely randomized trees (ERT) which is a generic algorithm and is applicable to the survival context, multiple trees are generated without resampling but by selecting a random subset of variables and one arbitrary cut-off point for each variable at each node. 
For a comprehensive review about survival trees and random survival forests, see \cite{bou2011}, \cite{rsfReview}, and \cite{zhou2015}.

One of the characteristic features of the tree-based methods in survival analysis is the way of incorporating censored information into measuring heterogeneity. While Classification And Regression Trees (CARTs, \cite{breiman1984}) and Random Forests \citep{breiman2001}, designed for continuous outcomes, use mean squared error (MSE) for quantifying heterogeneity, in right-censored survival tree methods, alternative approaches such as the log-rank statistic \citep{rsf} and inverse probability weighting \citep{molinaro2004, steingrimsson2019} are used. 

Using the log-rank statistic, however, can cause bias for two reasons. First, the log-rank statistic assumes that censoring time is independent of failure time. In practice, censoring is often informative of failure time. Thus, when the independent censoring assumption is violated, survival trees built based on the log-rank statistic may not be able to identify the optimal partition. Second, even when censoring is independent of failure, the log-rank statistic does not account for heterogeneity within each daughter node. In other words, the log-rank statistic implicitly assumes that subjects within a daughter node share the same marginal hazards process over censored intervals, when in fact, they may have different hazard processes conditional on their covariate values. This discrepancy contradicts and, as a result, possibly undermines the purpose of the random survival forests---estimation of the covariate-conditional hazards. Thus, naive use of the log-rank statistics could incur significant bias by choosing sub-optimal partitions.


\cite{rist} provided an intuitive solution to this problem by proposing recursively imputed survival trees (RIST) for right-censored data. The main idea is to guess the censored failure time using conditional survival probabilities and to utilize it for splitting. Considering that the finest covariate-conditional survival probabilities are available only after the trees grow far enough towards their terminal nodes, they use a recursion technique so that the terminal node prediction is utilized to impute the censored subjects in the next iteration of the forest building process. 

This issue, however, has yet to be fully addressed in the interval censored data literature. Moreover, tree-based regression methods for interval censored data are sparse; there are only a few tree-based methods available. \cite{yin2002} and \cite{fu2017} developed tree models for interval censored data that use the likelihood ratio test and a modified log-rank test as a splitting criterion, respectively. \cite{yao2019} recently extended the work of \cite{fu2017} to an ensemble method. \cite{yang2021} proposed a survival tree method for current status data that applies the idea of censoring unbiased transformation \citep{steingrimsson2019}. However, these methods have the aforementioned limitation of insufficient usage of covariate-conditional information.

To respond to this issue, we propose a tree-based nonparametric regression method for interval censored survival data. The method uses a recursion strategy \citep{rist} which incorporates a self-consistency equation \citep{efron1967}. In addition, we address additional challenges inherent to interval censored data:  first, the self-consistency algorithm may not identify the global optimum for interval censored data, and second, the interval censored data are highly noisy. To overcome such additional concerns, the method is equipped with a convergence monitoring procedure over recursions, probabilistic provision of information rather than imputation, and smoothing along the time domain. The proposed method shows high prediction accuracy both on simulated data and on our illustrative examples of avalanche victims data \citep{haegeli2011, jewell2013} and national mortality data \citep{sorlie1995}. An \texttt{R} package \texttt{icrf} is available on CRAN.

The rest of this paper is organized as follows. In Section~\ref{s:setup}, we describe the data structure and modeling assumptions. In Section~\ref{s:method}, the proposed methods are introduced and discussed in context. 
The uniform consistency of the method is derived in Section~\ref{s:consistency}. The predictive accuracy of the proposed method is evaluated using simulations and analysis of two sets of data in Sections~\ref{s:sim} and \ref{s:analysis}, respectively. In Section~\ref{s:discuss}, we discuss limitations and future areas for research.

\section{Data setup and model}
\label{s:setup}
The proposed method is applicable to interval censored data that include right-censored and current status data as special cases. 
Current status data, also also known as case-I censoring, only include survival status of a subject inspected at a single random monitoring time. The event time, $T$, is only known to lie within an interval $I \equiv (L, R]$, where $L=T^-$ and $R=T$ for an exactly observed $T$. Let $F(t)$, $F(t|X)$, $F(t|I)$, and $F(t|X, I)$ denote the marginal,  covariate-conditional, the interval-conditional, and the full-conditional distributions at time $ t$, respectively, where $X \equiv (X_1, ..., X_p) \in \mc X \subset\mb R^p$ is a $p$-dimensional covariate with distribution function $F_X(\cdot)$. We use $S \equiv 1 - F$ to represent a corresponding (conditional) survival function. 
For the censoring mechanism, we consider covariate-conditional non-informative censoring which is defined as, 
$$\Pr(T < t| L = l, R = r, L < T \le R, X) = \Pr(T < t| l < T \le r, X).$$
This implies that intervals do not provide any further information than the fact that the failure time lies in the interval given the covariate~\citep{oller2004, sun2007}.
The study length is denoted by $\tau < \infty$.
A random vector $U = (U_1, U_2, ..., U_M)$ denotes the monitoring times at each element of which the survival status of the subject is identified. $U$ follows a distribution $F_U$ with maximum potential number of follow-up times $M > 0$. Among the $M$ monitoring times, only one pair of two neighboring time points that includes $T$ contributes to the likelihood. Thus we only consider $\{L, R\} = \{U_{(m)}, U_{(m+1)}: U_{(m)} < T \le U_{(m+1)}, m = 0, 1, ..., M\}$ in the data analysis, where $U_{(m)}$ denotes the $m$th order statistic of the elements of $U$ with $U_0 \equiv 0$ and $U_{(M+1)} \equiv \infty$. Current status data correspond to $M=1$. 

\section{Interval censored recursive forests}
\label{s:method}

\subsection{Overview of the proposed method}

We adopt the recursion strategy for interval censoring and address the challenges of interval censoring---higher noise and non-identifiability of self-consistency algorithm---by carrying the full conditional survival probabilities of censored subjects, employing kernel smoothing of the survival curves along time, and monitoring convergence over recursion.

We outline the high level idea of the proposed method before we give a detailed description in the following subsections. As an initial step, to provide rough information about the censored intervals, we estimate the marginal survival curve, $S^{(0)}(t|X) = \hat S(t)$,
and obtain the estimate of the full conditional survival probability for each subject, $S(t|X_i, I_i)$ by projection. Instead of doing imputation as in RIST, we store the conditional probability information for each subject and use it in the splitting tests. In this way, we can avoid the Monte Carlo error resulting from the imputation procedures which can be significant for interval censored data. 
We develop the Generalized Wilcoxon's Rank Sum (GWRS) test and Generalized Log Rank (GLR) test that enable two-sample testing for interval censored data based on conditional probabilities. With one of those splitting rules selected, a predefined number of trees are built under a modified ERT algorithm. Unlike the original ERT algorithm, we subsample data to leave a small fraction (`the out-of-bag sample') of the data for later use.
At each terminal node of the trees, a local survival probability estimate is obtained in two ways: 1) the NPMLE of the survival curve is obtained based on raw interval data without using the survival curve information, or 2) the full conditional survival curves are averaged. We call the former a ``quasi-honest" approach, and the latter an ``exploitative" approach.
The tree survival probability estimates formed in this manner are averaged to obtain a forest survival probability estimate, $S^{(1)}(t|X)$, for the first iteration.
Then $S^{(k-1)}(t|X)$ is used to update the full conditional survival curve of each subject $S^{(k)}(t|X_i,I_i)$ at the $k$th iteration, $k=2, 3, ..., K$ . For each $k$, $\tilde S^{(k-1)}(t|X)$ is obtained by kernel-smoothing. The final prediction is then given by the smoothed survival curve at the iteration of the smallest out-of-bag error.
A detailed pseudo-algorithm is given in Algorithm~\ref{al:icrf}.

\begin{algorithm}[H]
\SetAlgoLined
\KwResult{$\tilde S(t|X) = \tilde S^{(k_{\text{opt}})}(t|X)$    where $k_{\text{opt}} = \arg\min_k \epsilon^{(k)}$\;}
 initialize $S^{(0)}(t|X)$ and kernel smooth ($\tilde S^{(0)}(t|X)$), if INITIAL\_SMOOTH is TRUE\;
 \For{$k ~(\text{forest iteration})=1,2,...,K$}{
  Update $S^{(k-1)}(t|X_i, I_i)$ based on $S^{(k-1)}(t|X_i)$ and $I_i$ for each $i$\;
  \For{$b ~(\text{tree construction})=1,2,...,K$}{
    Sample $\mb D_b$ of size $s = \lceil 0.95n\rceil$ from the dataset $\mb D$ ~ ($\mb D^{OOB}_b := \mb D \backslash \mb D_b$)\;
    Recursively partitioning using GWRS based on $\{S^{(k-1)}(t|X_i)\}$:\\
         At each node, randomly pick $\lceil \sqrt p \rceil$ variables, pick a random cut-off for each selected variable, 
             and find the optimal cut-off suggested by GWRS\;
     \eIf{QUASIHONEST}{
        $S_{b,l}^{(k)}(t|A_{b,l}) = $ NPMLE($\{I_i: X_i\in A_{b,l}\}$)\;
    } {
        $S_{b,l}^{(k)}(t|A_{b,l}) = \frac 1 {|A_{b,l}|} \sum_{X_i \in A_{b,l}}S^{(k-1)}(t|X_i, I_i)$\;
    }
    Kernel smoothing: $\tilde S_{b,l}^{(k)} = \text{KERNELSMOOTH}(S_{b,l}^k)$\;
    The conditional survival function for the tree:\\ 
    ~ $S_{b}^{(k)}(t|X) = \sum_{l=1}^{L_b} S_{b,l}^{(k)}(t|A_{b,l})1(X\in A_{b,l})$,
    ~ $\tilde S_{b}^{(k)}(t|X) = \sum_{l=1}^{L_b}\tilde S_{b,l}^{(k)}(t|A_{b,l})1(X\in A_{b,l})$\;
    The out-of-bag error for the tree: $\epsilon_b^{(k)} = IMSE(\tilde S_{b}^{(k)}, \mb D^{OOB}_b)$\;
  }
  Obtain the conditional survival function for the forest:\\
            $S^{(k)}(t|X) = \frac 1 {n_{\text{tree}}}\sum_{b=1}^{n_{\text{tree}}} S_{b}^{(k)}(t|X),$ ~ $\tilde S^{(k)}(t|X) = \frac 1 {n_{\text{tree}}}\sum_{b=1}^{n_{\text{tree}}}\tilde S_{b}^{(k)}(t|X)$\;
    Calculate the out-of-bag error for the forest: ~
            $\epsilon^{(k)} = \frac 1 B \sum_{b=1}^B \epsilon_b^{(k)}$\;

 }
 \caption{Pseudo-algorithm for ICRF}
 \label{al:icrf}
\end{algorithm}

\subsection{Splitting rules}

For right-censored data, \citet{peto1972} compared the two-sample test statistics including the Wilcoxon Rank Sum (WRS) test and the log-rank test. They showed that the log-rank test is the most locally powerful test under Lehman-type alternative hypotheses while WRS also has strong power under Log-normal mean-shift alternative hypotheses. Thus, these tests can be considered as potential splitting rules with some modifications for interval censoring.

For interval censored data, we develop two splitting rules by extending the WRS and log-rank tests. We also consider two existing score tests proposed by \citet{peto1972} that are used by existing tree-based methods \citep{fu2017, yao2019}.
Below we describe the four splitting rules and show the consistency property of the newly developed rules. Simulation results in Section~\ref{s:sim_split} show that our developed rules have on average better performance than existing alternatives.

\textbf{A. Generalized Wilcoxon's Rank Sum test (GWRS)}. The WRS test statistic, 
$$\tilde W_n = \frac 1 {n_1n_2}{\sum_{i\in G_1}\sum_{j\in G_2} \xi(T_{1,i}, T_{2,j})},$$ 
estimates $\tilde\theta = {\Pr}(T_{1} < T_{2}) + \frac 1 2 {\Pr}(T_{1}=T_{2})$ where $T_l$ is the survival time of a randomly chosen subject in group $G_l$,
$\xi(T_{1,i}, T_{2,j}) = 1(T_{1,i} < T_{2,j}) + \frac 1 2 1(T_{1,i} = T_{2,j})$, and $n_1$ and $n_2$ are the sample sizes of the two groups, respectively. The estimand can be alternatively expressed as $\tilde\theta (S) = 1 + \int_0^\infty \check S_{G_1}(t)dS_{G_2}(t),$ where $S_{G_l}(t) = \Pr(T_{l} > t|G_l), l=1,2,$ is the marginal survival probability of the $l$th group and $\check S(t) = \frac 1 2 S(t) + \frac 1 2 S(t^-)$ where half of the probability mass in the left continuity point is shifted toward the right.
In the presence of administrative censoring, $\mathring W_n = \frac 1 {n_1n_2}{\sum_{i\in G_1}\sum_{j\in G_2} \xi(\mathring T_{1,i}, \mathring T_{2,j})}$ estimates $\theta(S) = {\Pr}(\mathring T_{1,i} < \mathring T_{2,j}) + \frac 1 2 {\Pr}(\mathring T_{1,i}=\mathring T_{2,j}) = 1 + \int_0^{\tau} \check S_{G_1}(t)dS_{G_2}(t) - \frac 1 2 S_{G_1}(\tau)dS_{G_2}(\tau)$, where $\mathring T_{l,i} = T_{l,i}\wedge \tau, l=1,2$.

We then generalize this statistic to allow non-informative interval censoring as follows:
$$W_n (S) = \frac 1 {n_1n_2}{\sum_{i\in G_1}\sum_{j\in G_2} \zeta(I_{1,i}, I_{2,j}|X_{1,i}, X_{2,j}; S)},$$ 
where $\zeta(I_{1,i}, I_{2,j}|X_{1,i}, X_{2,j}; S) = \Pr(\mathring T_{1,i} < \mathring T_{2,j}| T_{1,i} \in I_{1,i}, T_{2,j}\in I_{2,j}, X_{1,i}, X_{2,j}; S) + \frac 1 2 \Pr(\mathring T_{1,i} = \mathring T_{2,j} | T_{1,i} \in I_{1,i}, T_{2,j}\in I_{2,j}, X_{1,i}, X_{2,j}; S)$. Note $\zeta(I_{1,i}, I_{2,j}|X_{1,i}, X_{2,j}; S) = 1 + \int_0^{\tau} \check S(t|I_{1,i},X_{1,i})$ $dS(t|I_{2,j},X_{2,j}) - \frac 1 2 S(\tau |I_{1,i},X_{1,i})S(\tau |I_{2,j},X_{2,j})$.

By the following theorem, the GWRS statistic, $W_n(S_n)$, is shown to be consistent for $\theta(S_0)$, for a sequence $S_n$ converging to the true survival function $S_0$. The proof of Theorem \ref{t:gwrs} is deferred to the Supplementary Materials and the consistency conditions of $S_n$ are provided in Theorem \ref{t:uc}.

\begin{theorem}
\label{t:gwrs}
For a fixed pair of sets $G_l\subset \mc X$ and
any sequence $S_n$ such that $\sup_{t\in[0,\tau], x\in\mc X} |S_n(t|x) - S_0(t|x)| \to 0$ in probability as $n\to \infty$, $W_n(S_n) \to \theta(S_0)$ in probability as $n\to\infty.$
\end{theorem}

\textbf{B. Generalized Log-Rank test (GLR)}.
The log-rank test statistic for uncensored data or right-censored data is given by
$$\widetilde {LR}_n = \frac{\sum_{j=1}^J \frac{Y_{2j}{D_{1j}}+ Y_{1j}{D_{2j}}}{Y_{\cdot j}}}
{\sqrt{\sum_{j=1}^J \frac{Y_{1j}Y_{2j}D_{\cdot j}( Y_{\cdot j}- D_{\cdot j})}{Y_{\cdot j}^2(Y_{\cdot j} - 1)}}},$$
where $J$ is the number of distinct observed time points, $Y_{l,j}$ and $D_{l, j}$  are the number of subjects at risk right before and the number of events at the $j$th time point in group $l$, respectively, for $l = 1, 2$; $Y_{\cdot j} = Y_{1,j} + Y_{2,j}$ and $D_{\cdot j} = D_{1,j} + D_{2,j}$.

Using the full-conditional survival probabilities $S_i(t) \equiv S(t|X_i, I_i)$, the log-rank test can be extended to a generalized log-rank test (GLR) for interval censored data:
$$LR_n(S) = \frac{\int_0^\tau \frac{Y_2(t;S){dN_1(t;S)}+ Y_1(t;S){dN_2(t;S)}}{Y(t;S)}}
{\sqrt{\int_0^\tau \frac{Y_1(t;S)Y_2(t;S)dN(t;S)( Y(t;S)- dN(t;S))}{Y(t;S)^3}}},$$
where $Y_l(t;S) = \frac{1}{n_l}\sum_{i\in{G_l}} S_i(t-), N_l(t) = 1 - \frac{1}{n_l}\sum_{i\in G_{n_l}} S_i(t), l = 1, 2$, $Y(t;S) = \lambda_{n,1} Y_1(t;S) + \lambda_{n,2} Y_2(t;S),$ $N(t;S) = \lambda_{n,1} N_1(t;S) + \lambda_{n,2} N_2(t;S),$ $\lambda_{n,l} = \frac{n_l}{n},$ and $n=n_1+n_2.$ Note that the statistic $LR_n(S)$ is $\sqrt {n\lambda_{n,1}\lambda_{n,2}}$ times smaller in scale than $\widetilde {LR}_n$.

The following theorem establishes consistency of the GLR for $$\rho(S_0) = -\frac{\int_0^\tau \frac {S_{0}(t-|G_2)dS_{0}(t|G_1) + S_{0}(t-|G_1)dS_{0}(t|G_2) }{S_{0}(t-|G_1\cup G_2)}}
{\sqrt{-\int_0^\tau \frac {S_{0}(t-|G_1)S_{0}(t-|G_2)S(t|G_1\cup G_2)dS(t|G_1\cup G_2)}{S_{0}^3(t-|G_1\cup G_2)}}}$$ for some disjoint subsets $G_l\subset \mc X, l = 1,2$, with the proof relegated to Supplementary Materials.

\begin{theorem}
\label{t:glr}
For a fixed pair of sets $G_l\subset \mc X$ and
any sequence $S_n$ such that $\sup_{t\in[0,\tau], x\in\mc X} |S_n(t|x) - S_0(t|x)| \to 0$ in probability as $n\to \infty$, $LR_n(S_n) \to \rho(S_0)$ in probability as $n\to\infty.$
\end{theorem}

\textbf{C. WRS-score test (SWRS)}.
\citet{peto1972} introduced asymptotic score statistics for interval censored data, one of which is the two sample WRS test. The test statistic is given by $\widetilde {SW}_n = \frac 1 {n_1}\sum_{i\in G_1} SW_{1,i} - \frac 1 {n_2}\sum_{i\in G_2}SW_{2,i},$ where
$SW_{l,i} = \hat S_{G_l}(L_{l,i}) + \hat S_{G_l}(R_{l,i}) - 1.$
To rely on the self-consistency scheme the test statistic is rewritten as 
$SW_n(S) = \frac 1 {n_1}\sum_{i\in G_1}SW_{1,i}(S) - \frac 1 {n_2}\sum_{i\in G_2}SW_{2,i}(S)$ with
$SW_{l,i}(S) = S(L_{l,i}|X_{l,i}) + S(R_{l,i}|X_{l,i}) - 1, ~ l=1,2.$

\textbf{D. Log-Rank-score test (SLR)}.
Another score statistic (SLR) based on the log-rank test was proposed by \citet{peto1972}. This statistic, under the self-consistency algorithm, can be written as $SLR_n(S) = \frac 1 {n_1}\sum_{i\in G_1} SLR_{1,i}(S) - \frac 1 {n_2}\sum_{i\in G_2}SLR_{2,i}(S),$ where
$$SLR_{l,i}(S) = \begin{cases}
\frac{S(L_{l,i}|X_{l,i}) \log S(L_{l,i}|X_{l,i}) - S(R_{l,i}|X_{l,i}) \log S(R_{l,i}|X_{l,i})}
{S(L_{l,i}|X_{l,i}) - S(R_{l,i}|X_{l,i})} & S(L_{l,i}|X_{l,i}) > S(R_{l,i}|X_{l,i}),\\
\log S(L_{l,i}|X_{l,i}) + 1 & S(L_{l,i}|X_{l,i}) = S(R_{l,i}|X_{l,i}).
\end{cases}$$

The best cut point is the one that maximizes $|W_n - \frac 1 2|$, $LR_n$, $|SW_n|$, or $|SLR_n|$. We use GWRS as our main splitting rule in the subsequent analyses. In Section~\ref{s:sim_split}, we illustrate how different splitting rules affect the prediction accuracy.



\subsection{Self-consistent random forest and convergence monitoring}

The proposed ICRF can be understood as a self-consistent estimator. The self-consistency algorithm \citep{efron1967} can be succinctly expressed as a solution to the equation $f(\cdot;\theta) = \mb P_n f(\cdot|\mc Z; \theta),$
where $\mb P_n$ is the empirical average operator with respect to random quantities denoted as script letters, $\mc Z$ is the observed data, and $f(\cdot;\theta)$ is a functional parameter of interest. For instance, the non-parametric maximum likelihood estimator (NPMLE) for interval censored data is a self-consistent estimator for the marginal survival probability that solves for $S$ in
$S(t) = \mb P_n S(t|\mc I),$
where $\mc I$ is the observed intervals.

This algorithm can also be extended to tree-based estimators for survival probabilities. 
Without the self-consistency scheme, survival forest estimators can generally be written as 
$$\hat S(t|x) = \frac 1 {n_{\text{tree}}} \sum_{b=1}^{n_{\text{tree}}}\mb P_n \Bigg[ S_{1}(t| A_b(x;S_{2}))\frac{1(\mc X \in A_b(x; S_{2}))}{|A_b(x; S_{2})|/n}\Bigg],$$
where $A_b(x)$ is the terminal node of the $b$th tree that contains $x$, $|A|$ is the sample size of node $A$, $S_{1}(\cdot|A_b)$ is the survival probability estimate of the terminal node $A_b$, $S_{2}$ is the survival probability that is used to support splitting decisions in trees, and the subscripts indicating the dependencies with the tree index $b$ and the sample index $n$ are suppressed in $S_{1}$ and $S_{2}$. Note that $S_{2}$ is needed for tree partitioning, only when the failure time is censored. If there is no censoring, survival forest estimators can be reduced to $\hat S(t|x) = \frac 1{n_{\text{tree}}} \sum_{b=1}^{n_{\text{tree}}}\mb P_n \Bigg[ S_{1}(t| A_b(x))\frac{1(\mc X \in A_b(x))}{|A_b(x)|/n}\Bigg]$. Without censoring, the self-consistency of random survival forests can be achieved under certain smoothness assumptions by replacing $\hat S$ and $S_{1}$ with $S$ and incorporating an appropriate splitting rule. Splitting rules bring consistency to tree or random forest estimators if every terminal node of the resulting tree partition has an arbitrarily small length in probability for every side that contains signal and at the same time has arbitrarily many sample points, as the sample size grows larger \citep{cui2017}. Random splitting \citep{wager2015, wager2018} is often used for theoretical purposes, as is assumed in Theorem \ref{t:uc}, instead of the greedy splitting rules \citep{breiman2001}. However, since our paper is primarily directed at heuristic approaches based on the self-consistency concept, we will retain use of our modified ERT algorithm for splitting in simulations and data analyses.

Different survival tree methods assume disparate $S_{2}$ in the literature. For example, the marginal survival probability estimate $\hat S(t)$ is used in \cite{fu2017} and \cite{yao2019} and a node marginal survival probability estimate $\hat S(t|A)$ is used in \cite{rsf} and \cite{yin2002}. 
Note that most existing tree-based survival estimators have three survival quantities, $\hat S(t|x), S_{1}$ and $S_{2}$, that do not coincide with each other and thus, they are not self-consistent. 
This discrepancy between survival probabilities may cause a greater bias. Splitting based on crude information, e.g., using the marginal survival probability estimate $\hat S(t)$ or the intermediate node survival probability estimate $\hat S(t|A)$ as $S_{2}$ rather than using $S(t|x)$, results in greater finite sample bias. 
See the discussion of \citet{cui2017}, where the authors discuss the bias of random survival forests for which the splitting rule is based on the candidate node marginal survival probabilities.

Self-consistency can be derived by replacing $\hat S(t|x), S_{1}$ and $S_{2}$ with $S$. The ICRF estimator $\hat S$ solves for $S$ in
$$S(t|x) = \frac 1 {n_{\text{tree}}} \sum_{b=1}^{n_{\text{tree}}} \mb P_n S(t|\mc I, \mc X \in A_b(x; S))\frac{1(\mc X \in A_b(x; S))}{|A_b(x; S)|/n}.\eqno{(3.5.1)}$$
The self-consistency equation can be solved by recursion. 
This self-consistent estimator makes sense when $S^{(k)}(t|x) \simeq S(t|x)$ for some large $k$. 
However, self-consistency equations in general may have multiple solutions (non-identifiability) and thus, recursion algorithms may not guarantee convergence to the truth; for example, This issue arises when estimating the NPMLE for interval censoring \citep{wellnerZhan}.
For some initial guesses, the estimator may give an inconsistent estimate. Thus, it is crucial to make sure that an additional forest iteration brings reduction in error. To monitor this in the absence of knowing the true survival curve, the out-of-bag samples are used for estimating the accuracy. That is, for each tree in ERT, we randomly subsample a large fraction, e.g. 95\%, for tree construction
and evaluate the tree using the small (5\%) hold-out sample. Using a metric that will be discussed in Section~\ref{s:measure}, we monitor the performance of the ERT's over a prespecified number of iterations, e.g. $K=10$.


\subsection{Quasi-honesty}
\label{s:honesty}
Once partitioning procedures are done, the terminal node survival curves are estimated either i) by applying NPMLE to the raw interval data (\emph{quasi-honest} prediction) or ii) by averaging the full conditional survival curves (\emph{exploitative} prediction). The former approach is quasi-honest, as the survival probability of the previous iteration is only used in the partitioning procedure but not in the prediction procedure. It is not genuine honesty  \citep{athey2016}, in the sense that ICRF still uses the same interval data in both partitioning and terminal node prediction.

The second approach is exploitative. This approach is computationally efficient, 
since the prediction does not require a complicated optimization procedures, it is computationally light.  However, as is discussed in the following paragraphs, this approach tends to have higher bias, non-convergence, and dilution of signals.
RIST, where imputed values containing the information about the covariate-conditional survival curve are used for both partitioning and terminal node prediciton, is hence exploitative.

The role of (quasi-) honesty in the prediction accuracy should be understood in terms of the bias-variance trade-off. While honesty induces higher variability by not utilizing the whole information at each procedure, it relaxes the overfitting problem and makes trees less biased by maintaining less dependence between the partitioning procedure and the terminal node prediction procedure. Hence, quasi-honesty may or may not be beneficial to interval censored survival analysis. A large amount of information about the true survival curve is lost due to interval censoring. 
This means that there might be a room for an exploitative approach to make up the information loss, since it more fully utilizes the information. However, it is also true that once the estimation moves in a wrong direction initially, then the exploitative approach may keep driving the estimation sequence in the wrong direction, while the quasi-honest approach may suffer less from such non-convergence. 

Another property of the exploitative approach is dilution of signal. When the initial survival probability starts with the marginal survival distribution, even after partitioning, two different points in a feature space share a significant amount of information about the survival distribution. This results in lower variance and hence, sometimes, underfitting. This exploitative approach should therefore be used when the features do not contain a large amount of information about the failure time distribution. 
We compare the performances of these two approaches in Sections~\ref{s:sim} and \ref{s:analysis}.

\subsection{Smoothed forests}

Random forests are relatively smoother than base learners with respect to features. However, they are still discrete in the time domain, especially for the NPMLE of interval censored data. Since in reality the survival function is unlikely to include step functions, it can be beneficial to assume some smoothness on the true survival function. \citet{groeneboom2010} proposed two ways of estimating smooth survival curves for current status data. Although their first method, the maximum smoothed likelihood estimator (MSLE), may not apply to general interval censored data, one can easily use the second method, the smoothed maximum likelihood estimator (SMLE), for such data. The idea is to find a non-smooth nonparametric maximum likelihood estimator (NPMLE), $\hat S(t)$, and use kernel smoothing to obtain an SMLE: $\tilde S(t) = 1 + \int_0^t \int_{\mb R_+} \frac 1 h k_h(s-u)d\hat S(u)ds$, where $k_h$ is a kernel function with bandwidth $h>0$. For survival forests, the SMLE is computed for each terminal node of each tree: $\tilde S^{k,b}(t|x) = \sum_{l=1}^{L_{k,b}}\tilde S_l^{k,b}(t|A^{k,b}_l)1(x\in A_l^{k,b})$, where 
$A_l^{k,b}$ is the $l$th terminal node in the $b$th tree of the $k$th forest iteration, $l = 1, 2, ..., L^{k,b}$, $b = 1, 2, ..., n_{\text{tree}}$, and $k=1,2,...,K$. Then the smoothed random survival forest is $\tilde S^k(t|x) = \sum_{b=1}^{n_{\text{tree}}}\tilde S^{k,b}(t|x)$.
In this paper we use a Gaussian kernel with bandwidth $h = c{n_{\min}}^{-1/5}$ where we choose $c$ to be the inter-quartile range of the marginal survival distribution estimate and $n_{\min}$ is the minimum size of the terminal nodes. For discussion on the choice of the bandwidth, see \citet{groeneboom2010}. For the boundary kernel near $t = 0$, we use a mirror kernel $\tilde k_h(t, u) = k_h(t, |u|)$ for $t \le 4h$. 

\section{Uniform Consistency of ICRF}
\label{s:consistency}

Although the recursion technique is intended for bias correction for finite samples, the large sample behaviour of ICRF is of interest. We present in Theorem \ref{t:uc} a uniform consistency result for the quasi-honest ICRF. The proof is provided in the Supplementary Materials. We only consider case-II censoring, since this result can be generalized without much difficulty to case-$K$ censoring for $K<\infty$ \citep{huang1997}.

\begin{assumption}[Absolutely continuous measure]
    \label{a:monitor}
    The probability measure of the failure time, $T$, is absolutely continuous with respect to that of the monitoring times $(L,R)$. Specifically, the joint density of the monitoring times is positive ($g(l, r|x) > 0$), if $0 < S_0(r|x) < S_0(l|x) < 1$, for all $x\in\mc X$, where $S_0$ is the true survival probability. 
\end{assumption}

\begin{assumption}[Lipschitz continuity of the failure and censoring survivor functions]
    \label{a:lipschitz}
    There exist constants $L_S$ and $L_G$ such that
    $|S_0(t\mid x_1)-S_0(t\mid x_2)|\le L_S \|x_1 - x_2\|_1$ and $|G(t_1, t_2\mid x_1)-G(t_1, t_2\mid x_2)|\le L_G \|x_1-x_2\|_1$ for all $x_1, x_2\in \mc X$ and $t, t_1, t_2\in[0,\tau]$, where $G$ is the censoring survival distribution and $g$ is its derivative with respect to time.
\end{assumption}

\begin{assumption}[Weakly dependent covariate values]
    \label{a:covariate}
    The covariate space $\mc X$ is a $p$-dimensional unit hypercube, i.e., $X\in \mc X = [0, 1]^{p}$. $X$ has a density $f_{X}$ such that $\zeta^{-1} \le f_{X} (x) \le \zeta$ for all $x\in\mc X$ and some constant $\zeta \ge 1$.
\end{assumption}

\begin{assumption}[$\alpha$-regular and random-split trees]
    \label{a:split}
    Trees in the ICRF are random-split and $\alpha$-regular according to Definitions 3 and 4 of \cite{wager2018}.
\end{assumption}

\begin{assumption}[Terminal node size]
    \label{a:terminal}
    The minimum size $n_{\min}$ of the terminal nodes  in the ICRF trees grows at the following rate:
    $$n_{\min} \asymp n^\beta, \frac 1 2 < \beta < 1,$$
    where $a \asymp b$ implies both $a = \mc O(b)$ and $b = \mc O(a)$.
\end{assumption}

\begin{theorem} [The uniform consistency of interval censored recursive forests]
\label{t:uc}
    Suppose Assumptions~\ref{a:monitor}--\ref{a:covariate} hold. Then the interval censored recursive forest $\hat S_n$ built based on Assumptions \ref{a:split}--\ref{a:terminal} and quasi-honesty is uniformly consistent. That is,
    $$\sup_{t\in[0,\tau], x\in\mc X} | \hat S_n(t\mid x) - S_0(t\mid x)|\to 0$$ in probability as $n\to \infty$.
\end{theorem}

\section{Simulations}
\label{s:sim}
In this section, we run simulations in order to evaluate the prediction accuracy of ICRF in multiple aspects. We also discuss the computational cost of the method.
The first set of simulations is to compare the prediction accuracy of ICRF to that of existing methods under multiple scenarios. The second set of simulations is to compare the performances of different splitting rules of ICRF and to compare the performances of quasi-honest and exploitative prediction rules. The final set shows the performance as sample size grows.

The competitors considered include the Cox proportional hazards model \citep{finkelstein1986} which is implemented using the R package \texttt{icenReg} \citep{icenReg}, the survival tree method for interval censored data \citep{fu2017}, and the survival forest method for interval censored data \citep{yao2019}.

All the models except the Cox model are implemented using an R package \texttt{icrf} (version 2.0.0). 
Note that since ICRF estimates are a weighted average of NPMLE's and the method of \cite{yao2019} provides an NPMLE of weighted individuals, implementation of the latter by \texttt{icrf} might involve finite sample differences. 
Because for other methods than ICRF, the estimates are not identifiable at each time point but are uniquely obtained only as a set of probability masses in intervals, we interpolate the within-interval survival curve assuming a uniform density within those intervals. However, when the length of the intervals is not finite, an exponential density is assumed. That is, given the estimated probability  $\hat p_{[a, \infty)}$ of the last unbounded interval, the interpolated survival estimate is given by $\hat S(t) = 1 - \hat p_{[a, \infty)}^{-t/a}$ for $a \le t < \infty$.
The NPMLE often assigns a probability mass to the last bounded time interval even when there are subjects known to have failure times in an unbounded interval. This can deflate survival curves in the tail drastically. Such non-regularity can be relaxed by posing structures to the estimator \citep{anderson2016, polyanskiy2020}. In those cases, a correction is made so that the last probability mass is allocated exponentially over an unbounded interval.
We further include smoothed versions of the existing methods for a fair comparison. 
The code for the simulations is provided in the Supplementary Material.

\subsection{Generative models and tuning parameters}
We first define the simulation settings by describing generative models and tuning parameters for the estimators. The basic framework for the generative models is largely taken from \citet{rist}.

\textbf{Generative models.} 
Six scenarios for two different monitoring times ($K=1$ and $K=3$) are studied. Scenario 1 (PH-L) assumes a proportional hazards model with linear hazards ratio, Scenario 2 (PH-NL) has a nonlinear hazards ratio (PH-NL) in place of Scenario 1, and the third (non-PH) is a non-proportional hazards model, where all three scenarios assume non-informative censoring. The fourth scenario (CNIC) has non-informative censoring conditional on $X$, and the fifth scenario (IC) has informative censoring. To further study how smoothed estimators behave under a non-smooth true survival curve, we further adopted Scenario 6 (non-SM), where the first scenario is modified so that the density of the event times is degenerate. The settings are defined more concretely in Table~\ref{tab:setting}. The sample size $n$ of the training sets is 300 and samples are independently drawn. The study period ($\tau$) is set to 5 for all scenarios.

\begin{table*}[ht]
    \centerline{
    \begin{tabular}{lcccccccccc}
         scenario & $X$ & $p$ & $T$ & $U_k$ & $\mu$  & $\rho$ & $var(\mu)$ \\
         \hline
         1 PH-L & $N_{25}(0, \Sigma(\rho))$ & 25 & $Exp(\mu)$ & $Exp(\bar\mu)$ & $e^{0.1\sum_{j=11}^{20} X_j - 0.1}$ & 0.9 & 1.27\\
         2 PH-NL & $U([0,1]^{10})$ & 10 & $Exp(\mu)$ & $U([0, \tau])$ & $\sin(\pi X_1) + 2|X_2 - \frac 1 2| + X_3^3$ &  - & 0.24\\
         3 non-PH & $N_{25}(0, \Sigma(\rho))$ & 25 & $G(\mu, 2)$ &
         $U([0,\frac{3}{2} \tau])$ & $0.5 + 0.3|\sum_{j=11}^{15} X_j|$ & 0.75 & 0.68\\
         4 CNIC & $N_{25}(0,\Sigma(\rho))$ & 25 & $LN(\mu)$ &$LN(0.8\mu)$ & $0.3|\sum_{j=1}^5 X_j| + 0.3|\sum_{j=21}^{25}X_j|$ & 0.75 & 0.41\\
         5 IC & $N_{10}(0,\Sigma(\rho))$ &  10 & $Exp(\mu)$ & $LN(T)$ & $2 expit(X_1+X_2+X_3)$& 0.2 & 0.46\\
         6 Non-SM & $N_{25}(0, \Sigma(\rho))$ & 25& $SDE(\mu)$ & $Exp(\bar\mu)$ & $e^{0.1\sum_{j=11}^{20} X_j - 0.1}$ & 0.9 & 1.27\\
         \hline
    \end{tabular}}
    \caption{Simulation settings; Independent samples of size $n = 300$, $X = (X_1, ..., X_P)$; $\Sigma(\rho) = \{\sigma_{ij}(\rho)\}, \sigma_{ij} = \rho^{|i-j|}$; 
    $U = (U_{(1)}, ..., U_{(K)})$  with $K = 1, 3$ and elements $U_k$ (conditionally) independent of each other;
    $N_P(\mu, \Sigma)$, the $p$-dimensional normal distribution with mean $\mu$ and variance $\Sigma$;  
    $LN(\mu)$, the log-normal with mean $\mu$ and variance 1; 
    $U(A)$, the uniform distribution over $A$;
    $Exp(\mu)$, the exponential distribution with mean $\mu$; $G(\mu, \theta)$, the Gamma distribution with  shape $\mu$ and scale $\theta$; 
    $SDE(\mu)$, the semi-discretized Exponential defined as $\frac 1 2 (Exp(\mu) + \frac 1 2 \lceil 2 Exp(\mu)\rceil)$; 
    $\bar\mu$, a constant near the sample average of the $\mu$'s.}
    \label{tab:setting}
\end{table*}

\textbf{Tuning parameters.} 
The tuning parameters for the tree-based methods are summarized in Table~\ref{tab:tune}. The minimum size of the terminal nodes is 6 for ensemble learners and 20 for the non-ensemble tree method. For ensemble learners, 300 trees are built by considering randomly chosen $\lceil \sqrt p \rceil$ candidate variables at each node. The default splitting rule for ICRF is set as GWRS and both quasi-honest and exploitative prediction are used for terminal node predictions. However, other splitting rules are also compared. The marginal survival probability estimates are used as the initial guess. As for smoothing, the bandwidths are chosen to be $h = cn^{-1/5}$ with $c = \frac 12 [\hat S^{-1}(0.25) - S^{-1}(0.75)]$.

\begin{table*}[]
    \centering
    \begin{tabular}{cccccccc}
         method & $n_{fold}$ & $n_{\text{tree}}$ & $mtry$ & $s$ & $replace$ &$n_{\min}$  \\
         \hline
         ICFR & 10 & 300 & $\lceil \sqrt p \rceil$ & $\lceil 0.95 n \rceil$ & no & 6\\
         Fu & - & 300 & $\lceil \sqrt p \rceil$ & $\lceil 0.632 n \rceil$ & yes & 6\\
         Yao & - & - & - & - & - & 20\\
         \hline
    \end{tabular}
    \caption{Tuning parameters for tree-based methods; Fu, the method of \cite{fu2017}; Yao, the method of \cite{yao2019};
    $n_{fold}$, the maximum number of iterations for ICFR;
    $n_{\text{tree}}$, the number of trees making up the random forests;
    $mtry$, the number of candidate features on which splitting tests are done at each node;
    $s$, the size of the random resample for a tree in random forests;
    $replace$, whether to resample with replacement or not;
    $n_{\min}$, the minimum number of observations in terminal nodes.}
    \label{tab:tune}
\end{table*}

\subsection{Prediction Accuracy}
\label{s:measure}
To assess the prediction accuracy of the estimators, we use integrated absolute error and supremum absolute error over the study period. They are defined as $\epsilon^{INT}(\hat S_n) = \int_0^\tau |S_0(t) - \hat S_n(t)|dt$ and
$\epsilon^{SUP}(\hat S_n) = \sup_{t\in[0,\tau]} |S_0(t) - \hat S_n(t)|$, respectively. These error measurements are obtainable only when the true survival curve $S_0$ is available. To measure the error in the absence of the true survival curve, we use the integrated mean squared errors type 1 ($\text{IMSE}_1$) and type 2 ($\text{IMSE}_2$) \citep{banerjee2016}. $\text{IMSE}_1$ is defined as the squared discrepancy of the estimate from the actual survival status averaged over the interval of the known survival status and then averaged over the sample. That is, $\text{IMSE}_1(\hat S_n| \mb D) = \frac 1 n \sum_{i=1}^n \frac 1 {\tau - (R_i \wedge \tau) + (L_i \wedge \tau) } \big\{\int_0^{L_i\wedge \tau} (1 - \hat S_n(t|X_i))^2dt + \int_{R_i}^{R_i \wedge \tau} \hat S_n(t|X_i)^2dt \big\}$, where 
$\mb D = \{(L_1, R_1, X_1), ..., (L_n, R_n, X_n)\}$ is the test set. This can be regarded as a modified integrated Brier score \citep{graf1999}. 

$\text{IMSE}_2$ is defined over the whole time domain up to the study length, where the discrepancy over the censored interval is calculated by the difference between the covariate-conditional survival curve and the full-conditional survival curve: 
$$\text{IMSE}_2(\hat S_n| \mb D) = \frac 1 n \sum_{i=1}^n \frac 1 \tau \int_0^\tau (\hat S_n(t|X_i, I_i) - \hat S_n(t|X_i))^2dt.$$

As mentioned in the previous section, $\text{IMSE}_1$ is used for convergence monitoring of ICFR, as it is a model-free measure. The out-of-bag samples are used as a test set for measuring $\text{IMSE}_1$. The error measurement for convergence monitoring is given by
$$\text{IMSE}_1^{ICFR} (\hat S_n| \mb D) =  \frac 1 {n_{\text{tree}}} \sum_{b=1}^{n_{\text{tree}}} \text{IMSE}_1 (\hat S_{n,b}| \mb D_b^{OOB}),$$
where $\mb D$ is the whole training data and $\mb D_b^{OOB}$ is the out-of-bag sample left for the $b$th tree.

\subsection{Simulation results}
\subsubsection{Comparison with other methods}
\begin{figure*}[ht]
    \centering
    \includegraphics[width = 1\textwidth]{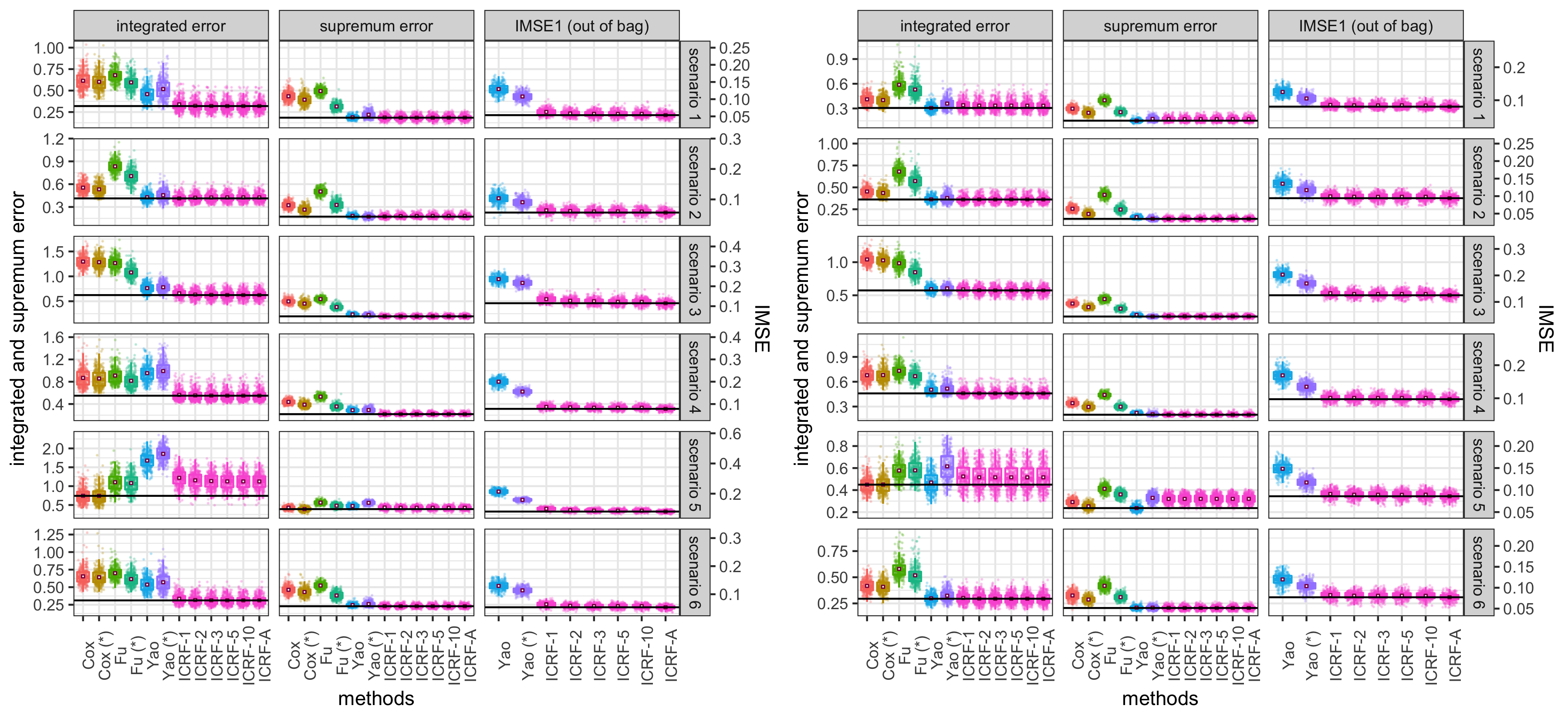}
    \caption{Prediction errors of methods under different simulation settings (the ICRF's are built in a quasi-honest manner); Fu, \cite{fu2017}; Yao, \cite{yao2019}; (*), smoothed versions;
    The boxes on the left column are for case-I censoring ($M=1$) and those on the right column are for case-II censoring ($M=3$); For each setting, the horizontal line indicates the minimum of mean error levels of the methods.}
    \label{fig:mainSim}
\end{figure*}

Simulations are done with $n_{sim} = 300$ replicates for each distinct setting. The simulation results based on quasi-honesty and GWRS rule are illustrated in Figure~\ref{fig:mainSim}. 
The results in the left column are for Case-I censoring and those in the right column are for Case-II censoring.
For convenience, we denote the ICRF estimator at the $k$th iteration by ICRF-$k$. The iteration with the best out-of-bag error among the ten iterations is denoted by $A$. In the results, ICRF-1, ICRF-2, ICRF-3, ICRF-5, ICRF-10, and ICRF-A are presented.

\textbf{Comparison with other methods}. For most of the scenarios, ICRF's have minimum or close-to-minimum integrated and supremum absolute errors. For Scenario 5 (both $M=1, 3$), where Cox models have better integrated absolute errors than the ICRF's, ICRF's have as good supremum errors as the Cox models. Also noting that simpler models such as \cite{fu2017}'s method and the Cox models have better accuracy than the method of \cite{yao2019} under Scenario 5, there is evidence that underfitting might be beneficial for settings where features contain weak signals, i.e., when $var(E(T|X))$ is low.

In Scenario 1, where data are generated under the proportional hazards model, ICRF's have better average accuracy than that of the Cox models. 
Although the Cox models eventually have higher accuracy for larger samples (see Figure~\ref{fig:sampSize}), the results indicate that ICRF methods have a relatively high prediction accuracy.

\textbf{Convergence monitoring}. The ICRF's error rate often becomes smaller as the number of iterations increases on average. Although in general it decreases, it often fluctuates and sometimes increases. However, ICRF-A, the ICRF at the best iteration of $\text{IMSE}_1$ measured against the out-of-bag samples, have integrated and supremum absolute errors close to the minimums most of the time.

\subsubsection{Splitting rules and quasi-honesty}
\label{s:sim_split}
\begin{figure*}[ht]
    \centering
    \includegraphics[width = \textwidth]{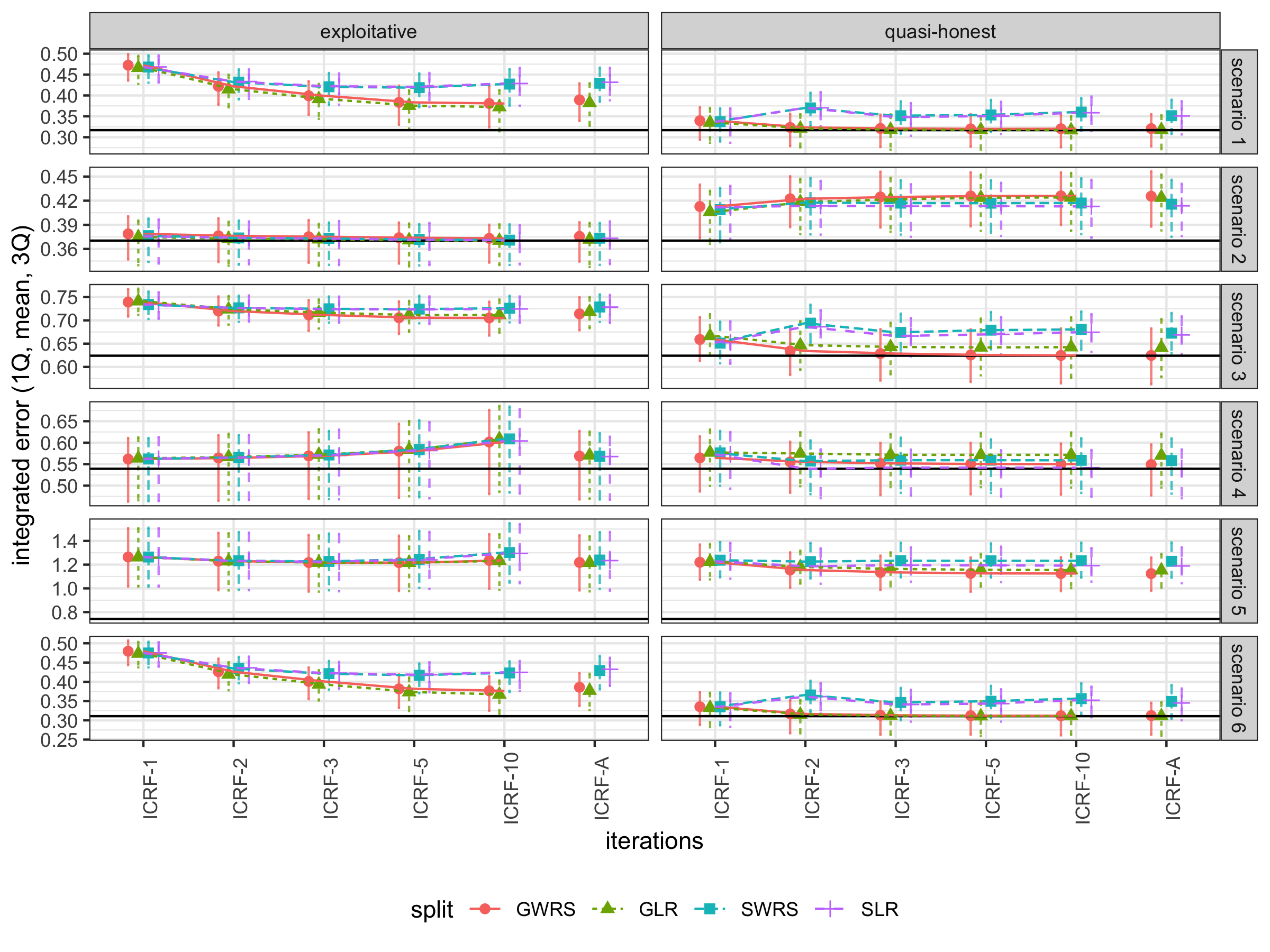}
    \caption{Mean and 1st and 3rd quartile $\epsilon^{INT}$ of splitting rules and prediction rules under Case-I censoring.}
    \label{fig:split}
\end{figure*}

Four splitting rules (GWRS, GLR, SWRS, SLR) with quasi-honest versus exploitative predictions are compared in Figure~\ref{fig:split} under six scenarios, with $M=1$ monitoring time. Most of the time, the new splitting rules (GWRS and GLR) have on average less error than the score-based rules (SWRS and SLR). Between GWRS and GLR, the two methods have about the same prediciton accuracy.  The gap between the new splitting rules and the score-based rules might reflect the fact that score-based rules rely on approximation, while GWRS and GLR do not.

On the other hand, the comparison between quasi-honest and exploitative predictions is less consistent. One does not always beat the other.
In Scenarios 2, the exploitative prediction has lower integrated absolute error, and in other scenarios, it has higher error rates. 
As mentioned in the last paragraph of Section~\ref{s:honesty}, exploitative prediction tends to make weak contrasts between two feature values and is expected to perform well when the true distribution has faint signals. In contrast, quasi-honest prediction provides more precise estimates when the signal is strong. As can be seen Figure A1 in the Supplementary Materials, exploitative prediction is computationally lighter than quasi-honest prediction and this difference overwhelms the difference made by different choices of the splitting rules. See the Supplementary Materials for more discussion on computational cost.

\subsubsection{Varying sample sizes}

\begin{figure*}[ht]
    \centering
    \includegraphics[width = \textwidth]{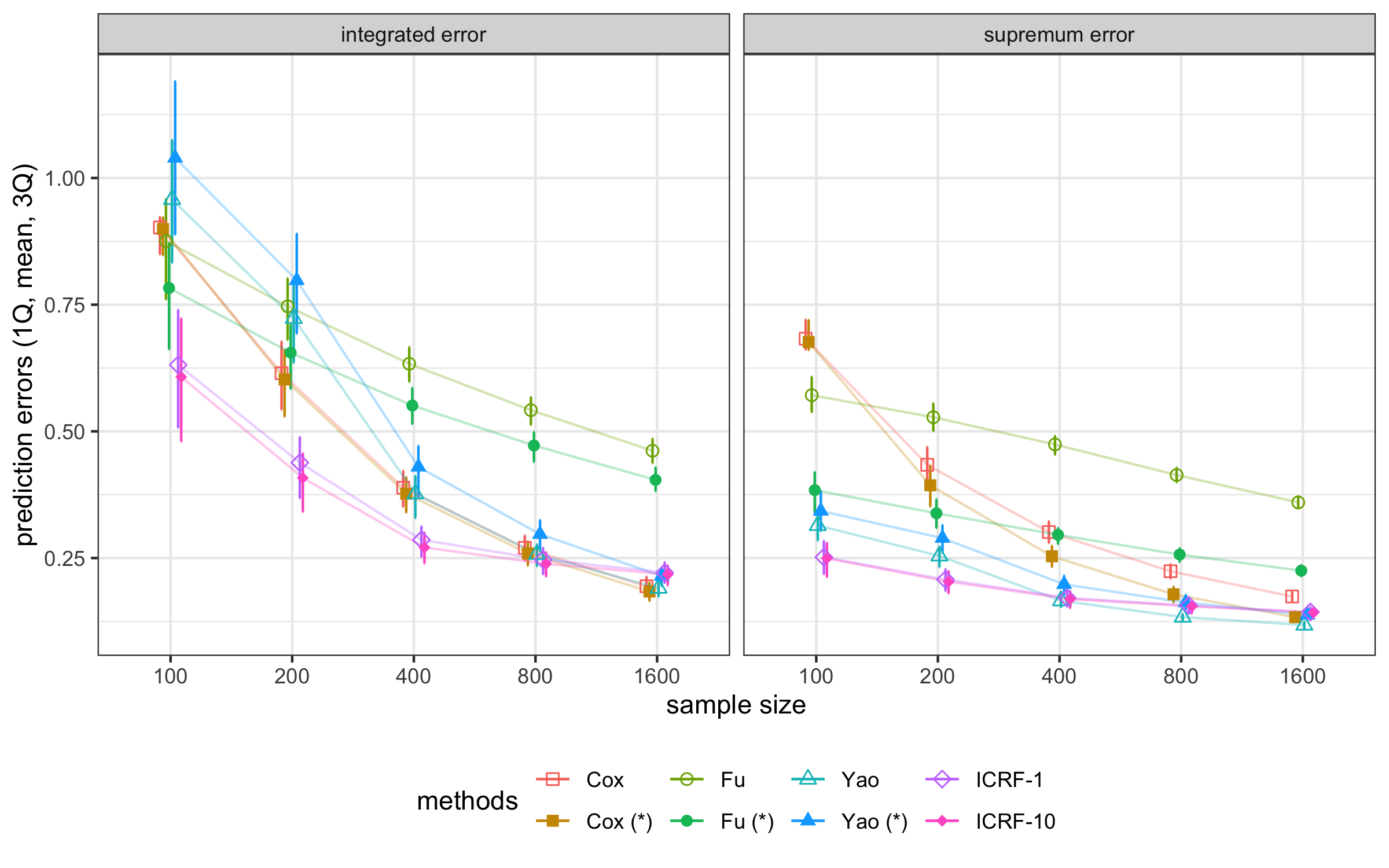}
    \caption{Prediction errors under different sample sizes for Scenario 1 and $K=1$.}
    \label{fig:sampSize}
\end{figure*}

The prediction accuracy of each method is evaluated under different sample sizes for current status data ($M=1$) under Scenario 1 (proportional hazards model). The integrated and supremum errors are measured. For ICRF, the last fold (10th) estimate is used for illustration. The mean, the 1st quartile, and the 3rd quartile of error measurements across 300 replicates are illustrated in Figure~\ref{fig:sampSize}.

The Cox model, although it does not have the smallest errors for small sample sizes ($n= 100, 200$), has rapidly decreasing errors as the sample size grows larger for both integrated and supremum absolute errors. Among the nonparametric models, ICRF shows the highest prediction accuracy in terms of all error measures for most sample sizes. For $n=1600$, the integrated error is lowest for the Cox model and has virtually a converged value for all ensemble-based methods.

The computation time of ICRF increases in a mildly superlinear fashion with respect to the sample size. See Figure A2 and Web Section 4 of the Supplementary Materials.

\section{Data analyses}
\label{s:analysis}
In this section, we apply ICRF (using 10 iterations), and three other methods---\cite{fu2017}, \cite{yao2019}, and the Cox model---with the corresponding smoothed versions to two existing data sets: (i) avalanche victim data where the time of discovery and a victim's survival status were only observed (\cite{jewell2013}), and (ii) data extracted from the National Longitudinal Mortality Study (\cite{sorlie1995}).

\subsection{The avalanche victims data}
\label{ss:avalanche}
The data of 1,247 avalanche victims buried in Switzerland and Canada between October 1980 and September 2005 are analyzed (\citep{jewell2013}). The dataset includes duration of burial and status of survival of the subjects, and thus can be regarded as current status data. The covariates include location, burial depth, and the type of outdoor activities involved. Approximately 10\% of the observations have missing burial depth. The main quantity of interest is the covariate-conditional survival probability where the event time is defined as time from burial to death. The event time here is counterfactual in a sense that the event time is the time until death had the person not been discovered (prior to death).

We use the following assumptions. First,  burial duration is independent of the time to event. This assumption is feasible as avalanche recovery is usually performed in the absence of knowledge of the survival status of victims. Second, the missingness of burial depth is completely at random. Although this assumption may not be fully valid, we analyze the data using complete cases only for comparative purposes.
Third, the survival of individual victims are independent. Since a single avalanche may involve multiple burials due to group activities, without a sufficient number of covariates, this assumption may not be valid. However, the point estimator of the survival function remains valid.

We randomly partition the complete data ($n = 1127)$ into training ($n = 789)$ and test ($n = 338$) datasets 300 times. The training sets are used for estimation of the survival curves, and the fitted models are evaluated using the corresponding test sets.
The avalanche data is highly skewed (median = 30, mean = 2,932, 3rd quartile = 110, max = 342,720 in minutes). To make the estimation computationally feasible, a log-transformed time domain is used with a transformation $h: [0, \infty) \mapsto [0, \infty)$ where $h(t) = \log(t + 1)$, and the prediction accuracy is evaluated in the transformed time domain. The study length is set as $\tau_t = 14400$ minutes (10 days) or $\tau = \log(\tau_t + 1) = 9.58$.  
The analyses are implemented using the \texttt{R} package \texttt{icrf} and the code is provided in the Supplementary Material. Preliminary parametric and semi-parametric regression analyses of the data are available in \citet{jewell2013}.

The prediction accuracy (IMSE) of the fitted models is summarized in Table~\ref{tab:avalanche} (LEFT). Among nonparametric methods, the ICRF with exploitative prediction has the best prediction accuracy. Although the smoothed Cox model shows the best prediction accuracy ($\text{IMSE}_1$ = 0.21, $\text{IMSE}_2$ = 0.19) among all available methods, the exploitative ICRF has a comparable performance ($\text{IMSE}_1$ = 0.22, $\text{IMSE}_2$ = 0.19).

\begin{table*}[ht]
\centering
\begin{tabular}{lcclcc}
  \hline
  \multicolumn{3}{c}{Prediction error} & \multicolumn{3}{c}{Variable importance}\\
 method & $\text{IMSE}_1$ (sd)  & $\text{IMSE}_2$ (sd) & variable & quasi-honest & exploitative\\ 
  \cmidrule(lr){1-3} \cmidrule(lr){4-6} 
  ICRF (Q) & 0.026 (0.0032) & 0.026 (0.0038) & \multicolumn{3}{c} {by $\text{IMSE}_1$ (multiplier = 0.0073)}\\ 
  ICRF (E) & 0.022 (0.0021) & 0.019 (0.0013) & Burial depth & 1.00 & 0.47\\ 
  Fu & 0.024 (0.0030)& 0.027 (0.0042) & Group activity & 0.17 & 0.41\\ 
  Fu (*) & 0.023 (0.0028) & 0.020 (0.0032) & Location &0.16 & 0.24\\ 
  \cmidrule(lr){4-6} 
  Yao & 0.025 (0.0031) & 0.026 (0.0031)& \multicolumn{3}{c} {by $\text{IMSE}_2$ (multiplier = 0.0041)}\\ 
  Yao (*) & 0.026 (0.0030) & 0.026 (0.0030)& Burial depth & 1.00 & 0.67\\ 
  Cox & 0.021 (0.0025) & 0.019 (0.0021)& Group activity & 0.27 & 0.46\\ 
  Cox (*) & 0.021 (0.0026) & 0.019 (0.0022) & Location &0.55 & 0.27\\ 
   \hline
\end{tabular}
\caption{Average prediction error of the avalanche survival models for each method (LEFT) and variable importance of the ICRF model fitted on the first training set of the avalanche data (RIGHT). ICRF (Q), quasi-honest ICRF; ICRF (E), exploitative ICRF;  The  importance  values  are  rescaled  so  that  maximum  values  for  each measure becomes 1.  The multiplier is the original importance scale.}
\label{tab:avalanche}
\end{table*}

\begin{figure*}
    \centering
    \includegraphics[width = \textwidth]{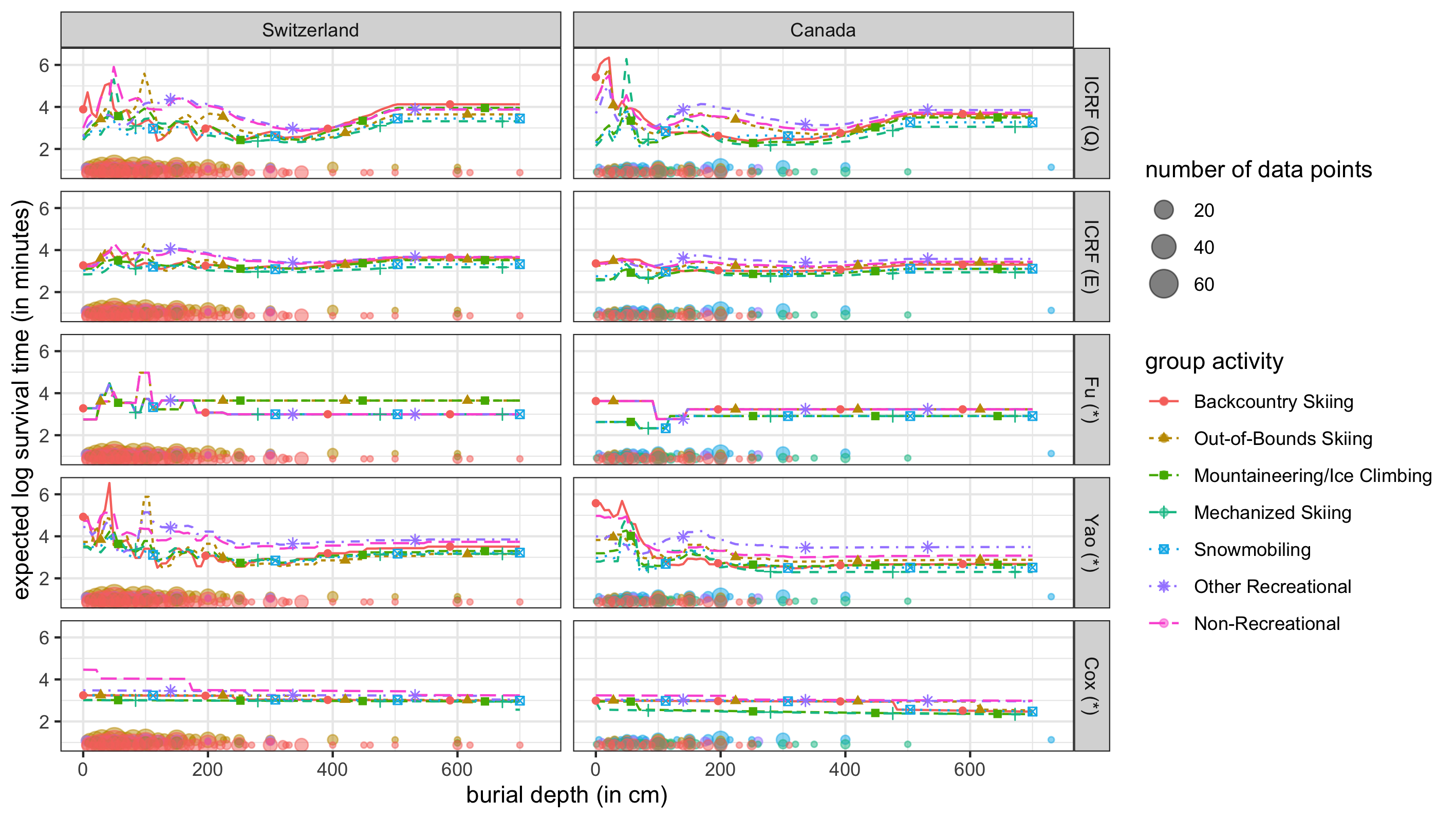}
    \caption{Estimated mean truncated log survival time in the avalanche data. The size of dots at the bottom of each box represents the number of sample data points.}
    \label{fig:avalET}
\end{figure*}

Figure~\ref{fig:avalET} illustrates the expected truncated log survival time, $\int_0^\tau h(t) dS(h(t)) + \tau S(\tau)$, of avalanche victims estimated by each smoothed model. While the Cox model, by assumption, has a monotone expected survival time with respect to each of the covariates, nonparametric models show non-monotone curves. The expected truncated survival time curves of the two prediction rules have a significant difference in their model variability, or $var[E[T|X]]$. Quasi-honest ICRF, compared to exploitative ICRF, has a wigglier curve along burial depths and has wider gaps among different group activities.

For most models, burial depth seems to be the most important covariate. In general, the mean truncated survival time decreases as the burial depth increases. However, for the emsemble methods (ICRF, \cite{yao2019}), the mean survival time increases for depths greater than 350 cm. This is considered to be an overfitting problem in a sparse data region.
In many models, the location also plays as important a role as burial depth; In the Cox model, the mean survival time in Canada is on average smaller than in Switzerland. Unlike the Cox model, nonparametric models have different patterns of expected survival time curves for different countries.

Variable importance is formally quantified by measuring the increase in $\text{IMSE}$ for a dataset where the values of each covariate in the original dataset are randomly permuted across the sample. The permutation is outside of the tree building procedure and does not affect the final prediction. The increase in IMSE is averaged across ten sets of random permutations. A larger increase in error for a variable indicates higher importance of the variable. The variable importance calculated for the model fitted on the first training set of the avalanche data is presented in Table~\ref{tab:avalanche} (RIGHT).
For either type of measurement ($\text{IMSE}_1$ or $\text{IMSE}_2$), burial depth is the most important variable explaining the survival probability. 
Group activity is chosen as more important than location except when importance is measured using $\text{IMSE}_2$ for the quasi-honest rule.

\subsection{National Longitudinal Mortality Study}
\label{ss:nlms}
We use the National Longitudinal Mortality Study (NLMS) data to explore the ability of the proposed method to model rich covariate information for survival data. The NLMS is a collaborative effort between the US Census Bureau and the National Heart, Lung, and Blood Institute (NHLBI), National Cancer Institute (NCI), National Institute on Aging (NIA), and the National Center for Health Statistics (NCHS). The views expressed in this paper are those of the authors and do not necessarily reflect the views of the Census Bureau, NHLBI, NCI, NIA, or NCHS. Among several data sets resulting from this extensive study, we use the dataset with six years of follow-up recorded around April 2002. The data are available at \url{https://biolincc.nhlbi.nih.gov/studies/nlms}.

The data include 0.7 million subjects with time to mortality, demographic information such as age, sex, and race, socioeconomic data such as income and housing tenure, and other covariates. Censoring is very high (97\% survived six years), as this is a general population sample, but only administrative censoring was observed. We narrow our focus to the elderly (age $\ge 80$ in years at entry) with complete covariate records ($n = 3,630$) and artificially induce current status censoring where the monitoring time depends only on age and household size. The proportion of missing covariate data is 20.7\% for the whole data and 65.9\% for the elderly subset. Thus, it should be noted that this data analysis is solely for performance comparison among the methods and that the results obtained from this regression analysis are limited to the selected population. The analysis framework is largely the same as for the avalanche data, except that with the increased sample size, the terminal node size was allowed to be larger ($n_{\min} = 20$ for random forests and $n_{\min} = 40$ for trees).
We provide further detail about the data, the pre-processing pipeline, and the censoring mechanism in the Supplementary Materials.

Table~\ref{tab:nlms} (LEFT) provides the prediction accuracy (IMSE) of the models trained and evaluated based on 70:30 cross-validation. The methods of \cite{yao2019} and the exploitative ICRF have similarly the lowest prediction errors among all methods including the Cox model. This indicates that strong assumptions such as proportional hazards and linearity may be violated in the data. Table~\ref{tab:nlms} (RIGHT) lists the variable importance according to ICRF. Besides age, type of health insurance (HI-type) turns out to be the most important variable that explains the failure time distribution, followed by presence of a social security number (SSN), self-reported health status (health), and sex. 

\begin{table*}[ht]
\centering
\begin{tabular}{lccccc}
  \hline
 \multicolumn{2}{c}{Prediction error} & \multicolumn{4}{c}{variable importance} \\ 
 method &$\text{IMSE}$ (sd) &  \multicolumn{2}{c}{quasi-honest} & \multicolumn{2}{c}{exploitative}\\
  \cmidrule(lr){1-2} \cmidrule(lr){3-6} 
  ICRF (Q) & 0.113 (0.0038)   & age      & 1.00 & age      & 1.00\\
  ICRF (E) & 0.113 (0.0065)   & HI-type  & 0.93 & HI-type  & 0.71\\ 
  Fu & 0.135 (0.0057)         & SSN      & 0.76 & SSN      & 0.54\\
  Fu (*) & 0.134 (0.0057)     & health   & 0.57 & health   & 0.45\\ 
  Yao & 0.112 (0.0042)        & sex      & 0.55 & sex      & 0.31\\ 
  Yao (*) & 0.111 (0.0038)    & race     & 0.20 & weight   & 0.24\\ 
  Cox & 0.117 (0.0055)        & tenure   & 0.15 & relationship & 0.18\\ 
  \cmidrule(lr){3-6}
  Cox (*) & 0.120 (0.0130)    & (multiplier) & 0.0169 & (multiplier) &  0.0151\\ 
   \hline
\end{tabular}
\caption{Average prediction error of the NLMS survival models for each method (LEFT) and variable importance of the ICRF model fitted on the first training set of the NLMS data based on $\text{IMSE}_1$ (RIGHT). For prediction error of the NLMS data, types 1 and 2 of the IMSE are equivalent. ICRF (Q), quasi-honest ICRF; ICRF (E), exploitative ICRF;  The  importance  values  are  rescaled  so  that  maximum  values  for  each measure becomes 1.  The multiplier is the original importance scale.}
\label{tab:nlms}
\end{table*}

\section{Discussion}
\label{s:discuss}
In this paper, we proposed a new tree-based iterative ensemble method for interval censored survival data. As interval censoring masks a huge amount of information, maximizing the use of available information can significantly improve the performance of estimators. Using an iterative fitting algorithm with convergence monitoring, ICRF solves the potential bias issue which most existing tree-based survival estimators have.
Specifically, this bias issue arises from not fully utilizing the covariate-conditional survival probabilities in the early phases of the tree partitioning procedure for these methods, which causes the kernel estimate to incur significant bias.
The WRS and log-rank tests were generalized for interval censored data and were used as splitting rules to fully utilize the hidden information. Quasi-honesty and exploitative rules were discussed for terminal node prediction. Smoothing adds another feature to ICRF.

We suggested many of the default modeling hyper-parameters, such as using GWRS or GLR as a splitting rule, the bandwidth of kernel smoothing, and the best iteration selection procedure by the out-of-bag $\text{IMSE}_1$ (or $\text{IMSE}_2$) measurement. However, the choice of the terminal node prediction rule remains unspecified. The quasi-honest and exploitative prediction rules each have their own strengths. The quasi-honest rule induces higher model variability, while the exploitative rule tends to favor simpler models. Thus, they perform well under high and weak signal settings, respectively. 

The challenge is that IMSE measurements are not always a good replacement for the true error measurement ($\epsilon^{INT}$ and $\epsilon^{SUP}$). The out-of-bag $\text{IMSE}_1$ measurement recommends the exploitative prediction rule for most of the simulation settings, including scenario 3 where the quasi-honest rule has higher accuracy than the exploitative rule.
Although the exploitative rule still beats the quasi-honest rule for five out of six scenarios and hence a decision rule based on out-of-bag $\text{IMSE}_1$ measurements may make sense, care must be taken.

This problem can be seen as a model selection problem balancing parsimony and flexibility. If the true model is thought to be smooth and simple, the exploitative rule should be employed. If the true model is believed to be complicated, the quasi-honest rule should be used. Unfortunately, the complexity or smoothness of true models is usually unknown. 
As model selection criteria such as AIC, BIC, and Mallow's Cp have been proposed in linear regression settings, new model selection criteria for interval censored survival models might greatly improve prediction accuracy.

The signal dilution property of the exploitative prediction rule might be caused by the fact that the marginal survival probability is shared by all censored subjects and the shared information is again carried forward to the next conditional survival probability estimate. This property might be mitigated by using non-marginal survival curves as the initial estimate. For example, the Cox model estimate or the first iteration of the quasi-honest ICRF estimate can be used as the initial estimate.

\if0\blind
{
\section*{Acknowledgements}
The authors thank Pascal Haegeli for providing the avalanche data \citep{haegeli2011} and the National Heart, Lung, and Blood Institute for sharing the NLMS data \citep{sorlie1995}. 
The third author was funded in part by grant P01 CA142538 from the National Cancer Institute.
}
\fi

\section*{Supplementary Materials}
 \begin{description}
      \item[R code:] R scripts which include simulations and data analyses in Sections \ref{s:sim}--\ref{s:analysis} (.zip file).
      \item[Proof: ] The proofs of Theorems 1--3.
      \item[Proof: ] Computational cost.
      \item[NLMS data analysis: ] More details about the NLMS data analysis.
 \end{description}

\bibliographystyle{plainnat}
\bibliography{ICRF.bib}

\label{lastpage}

\end{document}